\documentclass[aps,prl,twocolumn,groupedaddress]{revtex4-1}

\usepackage{graphicx}

\begin{document}

% Use the \preprint command to place your local institutional report
% number in the upper righthand corner of the title page in preprint mode.
% Multiple \preprint commands are allowed.
% Use the 'preprintnumbers' class option to override journal defaults
% to display numbers if necessary
%\preprint{}

%Title of paper
\title{Simple and efficient self-healing strategy for damaged complex networks}

% repeat the \author .. \affiliation  etc. as needed
% \email, \thanks, \homepage, \altaffiliation all apply to the current
% author. Explanatory text should go in the []'s, actual e-mail
% address or url should go in the {}'s for \email and \homepage.
% Please use the appropriate macro foreach each type of information

% \affiliation command applies to all authors since the last
% \affiliation command. The \affiliation command should follow the
% other information
% \affiliation can be followed by \email, \homepage, \thanks as well.
\author{Lazaros K. Gallos}
%\email[]{Your e-mail address}
%\homepage[]{Your web page}
%\thanks{}
%\altaffiliation{}
\affiliation{Department of Ecology, Evolution, and Natural Resources,
Rutgers University - New Brunswick, NJ 08901, USA}
\affiliation{DIMACS, Rutgers University - Piscataway, NJ 08854, USA}

\author{Nina H. Fefferman}
\affiliation{Department of Ecology, Evolution, and Natural Resources,
Rutgers University - New Brunswick, NJ 08901, USA}
\affiliation{DIMACS, Rutgers University - Piscataway, NJ 08854, USA}

%Collaboration name if desired (requires use of superscriptaddress
%option in \documentclass). \noaffiliation is required (may also be
%used with the \author command).
%\collaboration can be followed by \email, \homepage, \thanks as well.
%\collaboration{}
%\noaffiliation

\date{\today}

\begin{abstract}
The process of destroying a complex network through node removal has been
the subject of extensive interest and research. Node loss typically leaves
the network disintegrated into many small and isolated clusters. Here we
show that these clusters typically remain close to each other and we suggest a
simple algorithm that is able to reverse the inflicted damage by restoring
the network's functionality. After damage, each node decides independently
whether to create a new link depending on the fraction of neighbors it has
lost. In addition to relying only on local information, where nodes do not
need knowledge of the global network status, we impose the additional
constraint that new links should be as short as possible (i.e. that the new
edge completes a shortest possible new cycle). We demonstrate that this
self-healing method operates very efficiently, both in model and real
networks. For example, after removing the most connected airports in USA,
the self-healing algorithm re-joined almost 90\% of the surviving airports.
\end{abstract}

% insert suggested PACS numbers in braces on next line
% insert suggested keywords - APS authors don't need to do this
%\keywords{}

%\maketitle must follow title, authors, abstract, \pacs, and \keywords
\maketitle

\section{Introduction}

A property of critical importance for complex networks is their resilience
to damage or attack \cite{1,2,3,4,5}. One fascinating demonstration of the underlying
complexity of these systems is that compromise in structure can be
substantial even after the loss of a very small number of nodes \cite{6,7}. The
repercussions to the network from this structural compromise can be immense,
usually resulting in complete loss of communication among the surviving
nodes and therefore a complete destruction of the intended network
functionality. Because of the obvious importance for practical applications,
the robustness of a network's structure to damage has continuously remained
the focus of intensive research in the network science literature \cite{8, 8a, 8b, 8c, 8d, 8e}.

Even though we now have a thorough understanding of complex network
disintegration, the inverse process of `healing' a network is much less
understood. Is there a direct relationship between healing structural
features and restoring function? What does it really take for a network to
restore its functionality after losing some of its nodes? While there are
various types of healing, one realistic approach is to consider the case in
which disabled nodes have been permanently removed and cannot be
resurrected, but the surviving nodes can still generate new links to other
surviving nodes. Of course, a node could trivially create as many new links
as possible, but in many realistic settings the cost and time associated
with establishing a new link can be high. We then need to optimize the
network functionality given the constraints of rebuilding costs \cite{Schneider}. As a
natural first case, we explore the scenario in which optimal network
functionality is achieved under the structural condition that all the nodes
are connected in one large cluster. In this way, every node can reach any
other node in the network, even if the connecting path may be long. This
reflects systems such as communication networks \cite{9,10} electrical grids
\cite{11,12}, air traffic \cite{13} and shipping routes \cite{14}, etc. Additional
requirements can be imposed, for example the maximum path between any two
nodes can be bounded \cite{15}, or entirely different cases for structure and
function can be explored (e.g. function is optimized when the network
structure has a particular degree distribution, or is highly modular, etc.),
but as a first approach our only metric here is the size of the largest
cluster.

\subsection{The Duality of Structure and Function}

A basic, but fundamental insight is that healing can be employed with one of
two possible goals in mind. The first goal can be to fix the structure in a
way that the topology remains as close to the original network as possible.
For example, descriptions of some social groups have revealed particular
degree distributions and one natural inclination in healing a damaged social
network might be to restore edges by focusing on this aspect of repair.
However, degree distribution itself is unlikely to support social function.
For features such as social identity or support from a close group of
friends, the structural features of modularity and clustering are more
important. Therefore, if we are attempting repair in order to restore
function, we may decide to ignore entirely the impact to degree distribution
from repair, instead adding back links that restore (for example) local
clustering coefficient. Since damage can alter both structural and
functional features, it is easy to consider repairing structure since
regaining initial structure should provide restored initial function.
Critically, however, it may not be that one and only one structural feature
can support the original function. The main question is therefore to
discover efficient strategies that can restore network functionality through
the addition of new links. In a recent work, redundant (dormant) links were
considered to be activated after damage and it was shown that they can
restore functionality in infrastructure trees \cite{16,17}. Similarly, healing in
interdependent networks was shown to prevent cascading failures \cite{18}. In a
different approach, nodes were allowed to spontaneously recover and become
active again, leading to an interesting behavior of phase-flipping where the
network switches between high-activity and low-activity modes \cite{19,20}. Under
different definitions of network function in which a global minimum traffic
flow must be maintained and each node has a particular flow capacity that
must be redistributed in case of damage, the network functionality can be
restored by mitigation strategies \cite{21,22}. 

One critical and under-discussed feature of damage, especially in
inhomogeneous complex networks, is that the impact from the loss of a single
link on overall function can differ drastically depending on where that link
occurs in the local and global structure. Loss of a single link that
uniquely connects two otherwise disconnected components of a network will
have a greater impact to functions (such as the one we study) that rely on
the size of the largest cluster than would the loss of one of the three
links that make up a triangle. While obvious, it is nevertheless important
and profound that we consider healing as a process that should focus on
areas of the network in which the most damage has been done to the function,
rather than to the structure. In the case of our largest connected component
example, this means we should focus more effort towards creating new edges
that restore connectivity to nodes that are most direly impacted in their
ability to relay messages. 

\subsection{Local Repair for Global Function}

Phrased in this way, it may seem as though the only option is to analyze
global network structure in order to understand which nodes are most
critical to restoring function. Such a global algorithm could potentially
locate the optimal solution by mapping the current state of the network and
adding only the links that are missing to restore functionality. However,
this centralized planning may not always be feasible, since it may cost a
lot in terms of advance planning, communication between the central
authority and the individual nodes, the time that it takes to transmit all
relevant information, the possibility that communication takes place through
the network itself and the central node may have been functionally
disconnected from this critical communication, as well as any combination of
the above factors which may limit the ability of constructing one central
plan and communicate it to all the surviving nodes. Luckily, there is no
need for this type of centralized global analysis. We proceed to propose a
self-organizing healing algorithm that exploits this heterogeneity feature
of damage while relying only on local information accessible to each node,
and with low cost from the construction of new links. 

Since one of our goals will be to limit the amount of centralized
information that will be necessary before repair to the network can begin,
we propose purely local definitions of damage that can be assessed by each
node. To make this distinction clearly, in this work optimal `healing' will
refer to adding new links so that every node can reach any other node in the
system, and the term `self-healing' indicates that individual nodes decide
on their own whether they need to create new links or not.

\subsection{Self-Healing Algorithm to Restore Function}

We here demonstrate that our self-healing model can restore connectivity
function very efficiently in two different ways. Firstly, we show effective
restoration of function with the creation of very few new links relative to
the number lost to damage. We demonstrate the effectiveness of the method
through the size of the healed largest component, the number of nodes that
need to form new links, and the changes in modularity compared to the
original network. We further show this is true even if we impose cost
constraints on the length of new links (i.e. the number of links in the
shortest path required to connect the two nodes without the addition of the
new link). Secondly, we show that a null healing model (in which the same
number of links are constructed randomly) achieves a drastically lesser
restoration of function. We analyze the efficiency of our self-healing
algorithm over various network topologies and in cases of both
random-node-removal damage and when damage is targeted to affect only
specific, high-structural-impact nodes. Based on this `uneven damage to
function' perspective, we show that relatively inexpensive, rapid healing
may be relatively easy to achieve, even in the absence of global
information. 

To formalize the problem, we consider the simplest possible case of an
isolated complex network which undergoes a loss of a fraction of its nodes,
resulting in a possible loss of large-scale connectivity. To mitigate this
loss of function, each node can then generate one new link in an attempt to
restore connectivity. As mentioned above, this is a trivial problem without
any additional constraints because connecting to random nodes would result
in a structure similar to an Erdos-Renyi network. In our work, we study what
conditions are necessary and sufficient to restore connectivity under the
constraints that: a) establishing a new link is costly and that cost is
borne by the node initiating the link, b) the new links are as short as
possible (based on the path distances of the undamaged network), and c) the
decision is purely local, so that a node does not know anything about the
network state, except for its own neighborhood. We show that these
conditions can be easily met through a simple local-decision algorithm,
based on the number of surviving neighbors. There is no need to transfer any
information between nodes, and the only requirement is that a node can know
if a neighbor within a given distance remains alive or not.

\section{Results}

\subsection{Impact of node removal}

\begin{figure}
\includegraphics[width=.48\textwidth]{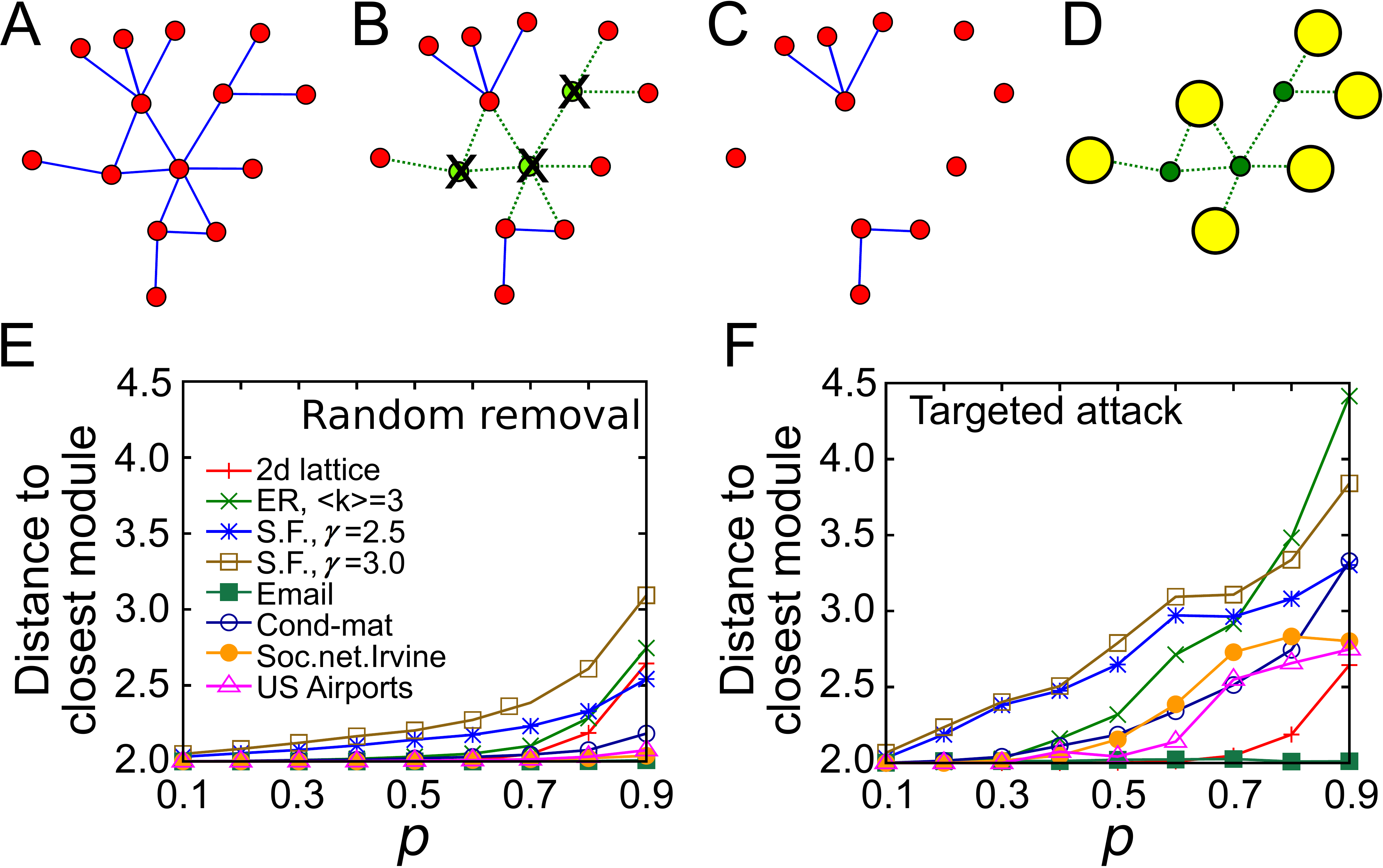}
\caption{\label{fig1} (Color online) Cluster distance after damage. (A) A toy network.
(B) A number of nodes are removed from the network.
(C) The remaining clusters in the network. (D) We consider each cluster to be one unit.
The distance between the clusters, $r_{\rm sep}$, is measured according to the distances
of cluster nodes in the original network. (E) The change of inter-cluster distance, $r_{\rm sep}$,
as a function of removed nodes, $p$, for model and real networks after random node removal.
(F) The change of inter-cluster distance, $r_{\rm sep}$, as a function of removed nodes, $p$,
for model and real networks after targeted node removal.}
\end{figure}

The first step in identifying efficient healing algorithms is to determine
the state of the network immediately after the removal process. This topic
has not received attention in the existing literature, which instead mainly
focuses only on the properties of the largest connected cluster. For the
purpose of a healing algorithm, we consider all the clusters that remain in
the system after a fraction $p$ of system nodes have been removed. A possible
measure that can quantify the extent of damage is the minimum distance
between these clusters, $r_{\rm sep}$, which we define as the minimum possible
distance from a node in the cluster to any other node in any other cluster.
Since all clusters are disconnected, this distance is calculated in terms of
the original network. In practice, we renormalize the damaged structure in
the following way (Fig.~\ref{fig1}A-D): Every cluster is substituted by one
super-node. All these super-nodes are obviously isolated. We restore all the
removed nodes and links among them. The links that were connecting removed
nodes with any node contributing to a super-node result in a link between
the super-node and the removed node (we also remove any double links). In
this renormalized network, we can then calculate how close the super-nodes
are to each other. Obviously, this distance is bounded by the minimum value
of $r_{\rm sep}=2$. Notice also that this distance is different from the distance
between a random node in the cluster and other clusters, since it is solely
determined by the one node in the cluster which is closest to another
cluster.

Intuitively, the inter-cluster distance is small in centralized networks and
is large in sparse, extended networks. For example, in a star network where
all nodes are connected to a central node, the distance $r_{\rm sep}$ can never be
larger than 2 independent of how many nodes we remove. In a one-dimensional
lattice this distance can become very large, especially when $p$ becomes large
and the gap among surviving clusters increases rapidly. Our focus on
inter-cluster distance is to enable estimation of the most efficient length
for the healing links. If, for example, typical inter-cluster distances were
$r_{\rm sep}=5$ then it would be pointless to add links much shorter than that.
Similarly, since one of our goals is to minimize the cost, calculating $r_{\rm sep}$
allows us to avoid constructing unnecessarily long links.

We studied the two typical cases of node removal in complex networks: random
removal and targeted attack. In Fig.~\ref{fig1}E we show the average $\langle r_{\rm sep} \rangle$ over all
clusters as a function of the fraction of randomly removed nodes, $p$. In
model networks, such as the two-dimensional square lattice, Erdos-Renyi, and
scale-free networks, the value of $\rangle r_{\rm sep}\langle$ remains close to 2 until roughly
$p=0.5$. When we remove a larger percentage of the nodes the inter-cluster
distance increases up to $\langle r_{\rm sep}\rangle \sim 2.5-3$. In real networks, on the other hand,
the average distance never increases significantly, independently of the
value of $p$.

In targeted attacks we remove the nodes in decreasing order of their degree.
As expected, the targeted attack leads to increasing damage and consequently
the inter-cluster distance increases in most of the networks (Fig.~\ref{fig1}F), with
the exceptions of the email network and the square lattice. The email
network includes many strong hubs and, as a consequence, node distances are
very short, which also reflects on the inter-cluster distances. For
lattices, a targeted attack is the same as random removal, because the
degree distribution is a delta function. Since ER networks also have a
narrow degree distribution, one might expect that the inter-cluster distance
in ER networks would also be independent of the attack strategy. In Fig.~\ref{fig1},
we find that the opposite is true. The average distance $\langle r_{\rm sep}\rangle$ increases
rapidly after $p=0.5$, and at $p=0.9$ it reaches the highest value we observed,
$\langle r_{\rm sep}\rangle\sim 4.5$. This is a result of the absence of hubs. When the majority of
the highest-connected nodes have been removed, the small remaining clusters
consist only of low degree nodes, which have a low probability of being
close to each other since there are no hubs to centralize the network.

So, in general, we observe that the isolated clusters are relatively close
to each other, and the inter-cluster distances remain for the largest part
close to the minimum value of $\langle r_{\rm sep}\rangle=2$, at least for random removals, while
in very few cases does it reach $\langle r_{\rm sep}\rangle=3$ or more under targeted attacks.

\subsection{Description of the method}

The basic idea of the self-healing method is that each node monitors the
fraction, $q$, of its neighbors that remain alive. If this fraction falls
below a given threshold $q_c$, e.g. $q_c$=50\%, then the neighbor attempts to
establish a new link. This link can be at a maximum distance of $r_{\rm max}$, where
distances are always measured as path length in the original network. If
there is no node within this distance, then the node abandons its attempt to
build a new connection. Importantly, in agreement with the above findings on
inter-cluster distance, we show that even if the maximum distance is the
smallest possible, i.e. $r_{\rm max}$=2, large-scale connectivity is still restored
even though the node has no information about whether or not the new link
will connect to the largest cluster, or even if it will connect to a
different cluster than the one it currently belongs to.

This process can be dynamic, so that nodes do not even need to be aware that
there is an attack, as long as they can monitor how many of their links lead
to live nodes, in which case they can respond immediately and attempt to
create a new link. This real-time healing is more effective than if the
nodes can only attempt to heal once an attack is completed (e.g. surprise
attacks, or cases in which new link construction is risky until an attack
has concluded) and obviates the need for rigorous definitions of global
attacks vs local damage. Here, we show that the self-healing algorithm is
efficient, even in this worst-case scenario, where the nodes do not have the
capacity to respond until the attack is over.

\begin{figure}
\includegraphics[width=.48\textwidth]{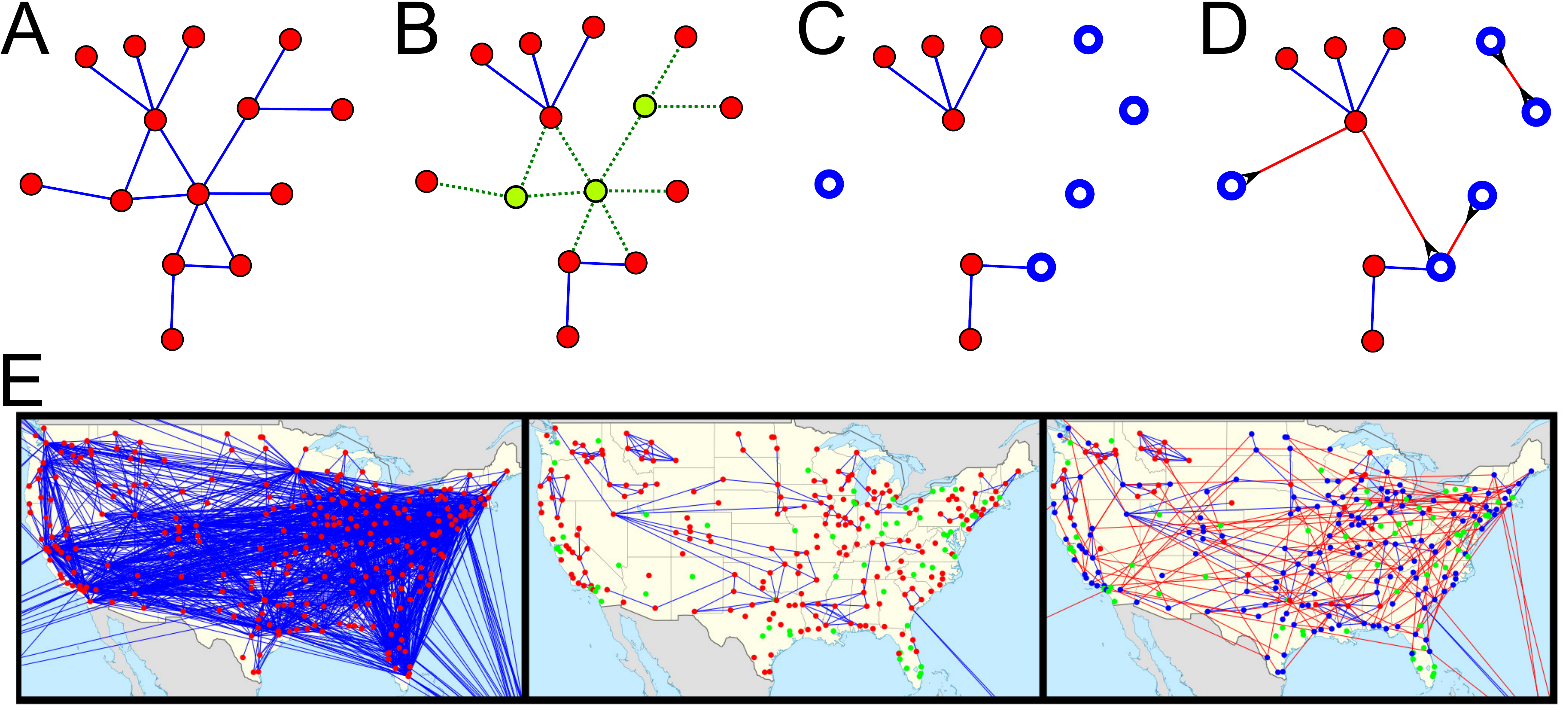}
\caption{\label{fig2} (Color online) Demonstration of the self-healing algorithm. (A) The initial structure
of a toy network. (B) Three nodes, indicated by green (light gray), are removed from the network along
with all their links. (C) The nodes which have lost half or more of their neighbors (blue nodes, empty symbols)
will seek new connections. (D) These nodes will choose to connect to a random surviving node,
if its distance $r$ in the original A network is at most $r_{\rm max}$. Here we use $r_{\rm max}=2$.
(E) (Left panel) Network of airport connections in USA. (Middle panel) A targeted attack removes
the 20\% most connected airport nodes, and results to a disconnected network. The removed nodes are shown
in green (light gray). (Right panel) The nodes that have lost more than 50\% of their neighbors (shown in blue, dark gray)
try to establish new links within distance $r_{\rm max}=2$. The new links (red lines) restore the large-scale network connectivity. }
\end{figure}

The order parameter, which quantifies the large-scale connectivity, is the
fraction of the live nodes that belong in the largest connected component,
denoted by $P_1(p)$ after the attack, and by $P_2(p)$ after the healing process.
The implementation of this model includes three steps (Fig.~\ref{fig2}A-D), which are
as follows:

\begin{enumerate}
\item A fraction $p$ of the $N$ nodes initially in the system is removed from
the network. The size of the remaining largest connected cluster, $P_1(p)$, is
expressed as a fraction of the remaining nodes.
\item Each of the remaining nodes decides independently if it needs a new
link, depending on whether the fraction of its remaining neighbors
$q=k_{\rm dam}/k_{\rm orig}$ exceeds a given threshold $q_c$ (for any given node, $k_{\rm dam}$ is the
number of neighbors after damage, and $k_{\rm orig}$ is the original number of
neighbors of this node). We denote the fraction of the surviving nodes that
need healing, i.e. those whose degree has fallen below the threshold, as $f$.
\item These $fN(1-p)$ nodes attempt to find a new neighbor within a distance
$r_{\rm max}$. If such a neighbor is found, the nodes establish the new link. The
fraction of nodes that succeed to establish a new link is $f_s\leq f$. These new
links then result to a new large cluster that includes a fraction of
$P2(p,q_c)$ [$\geq P_1(p)$] nodes.
\end{enumerate}

The size of the largest cluster, $P_1(p)$, indicates the extent of damage to
the structure immediately after the nodes removal. A value close to $P_1(p)=0$
means that the remaining nodes are isolated in small clusters and they
cannot access each other. The plot of $P_1(p)$ as a function of $p$ represents
the well-studied percolation process of the largest cluster as we increase
the number of removed nodes. The plot of $P_2(p,q_c)$ can be used to estimate
the efficiency of the healing process. After the self-healing process, the
largest cluster becomes $P_2(p,q_c)\geq P_1(p)$, since we have added links in the
structure. The difference between the two sizes is a measure of the process
success.

\subsection{Self-Healing of the Airport Network}

As a real-world example, we can demonstrate how the airport network in USA
can be influenced by attacks and healing (Fig.~\ref{fig2}E). The unperturbed network
(left panel) is quite dense and contains a large number of connections,
condensed in many hub nodes. After an intentional attack on 20\% of the most
connected nodes (middle panel), 80\% of the nodes are still functional but
are barely connected to other nodes, since most of the traffic was directed
through the hubs. The original network contains 332 nodes, so after the
removal of 66 nodes we are left with 266 nodes. The largest cluster size
connects only 30 of these nodes ($P_1\sim 15\%$), and the surviving nodes are either
completely isolated or belong to very small clusters. After we apply the
self-healing algorithm (right panel), the connectivity is restored and the
largest cluster grows to 235 nodes ($P_2\sim 88\%$). Of course, the network becomes
less centralized as hubs are removed and local connections are added during
healing, so that the network topology changes, but the important part is
that the functionality is restored and the network becomes navigable again.
Now, it is possible to reach practically any node, independently of the
origin. In this example it is also easy to see that the new connections have
to remain close to the healing node, since the cost of long-range
connections may be prohibitively high.

\subsection{Self-Healing in Model Networks}

The most interesting cases are those where the damage destroys the
large-scale connectivity (i.e. $P_1(p)$ tends to 0). The question now becomes
how easy it is to bring $P_2(p,q_c)$ as close to 1 as possible. Therefore, the
main quantities to compare are the values of $P_2$ vs $P_1$. An ideal process
would correspond to a plot of values as close to the horizontal $y=1$ line as
possible. This means that, independent of the initial damage, the
self-healing process guarantees complete connectivity. The worst-case
scenario, on the other hand, is a $y=x$ line along the diagonal, indicating
that there is no benefit from the self-healing process.

\begin{figure}
\includegraphics[width=.48\textwidth]{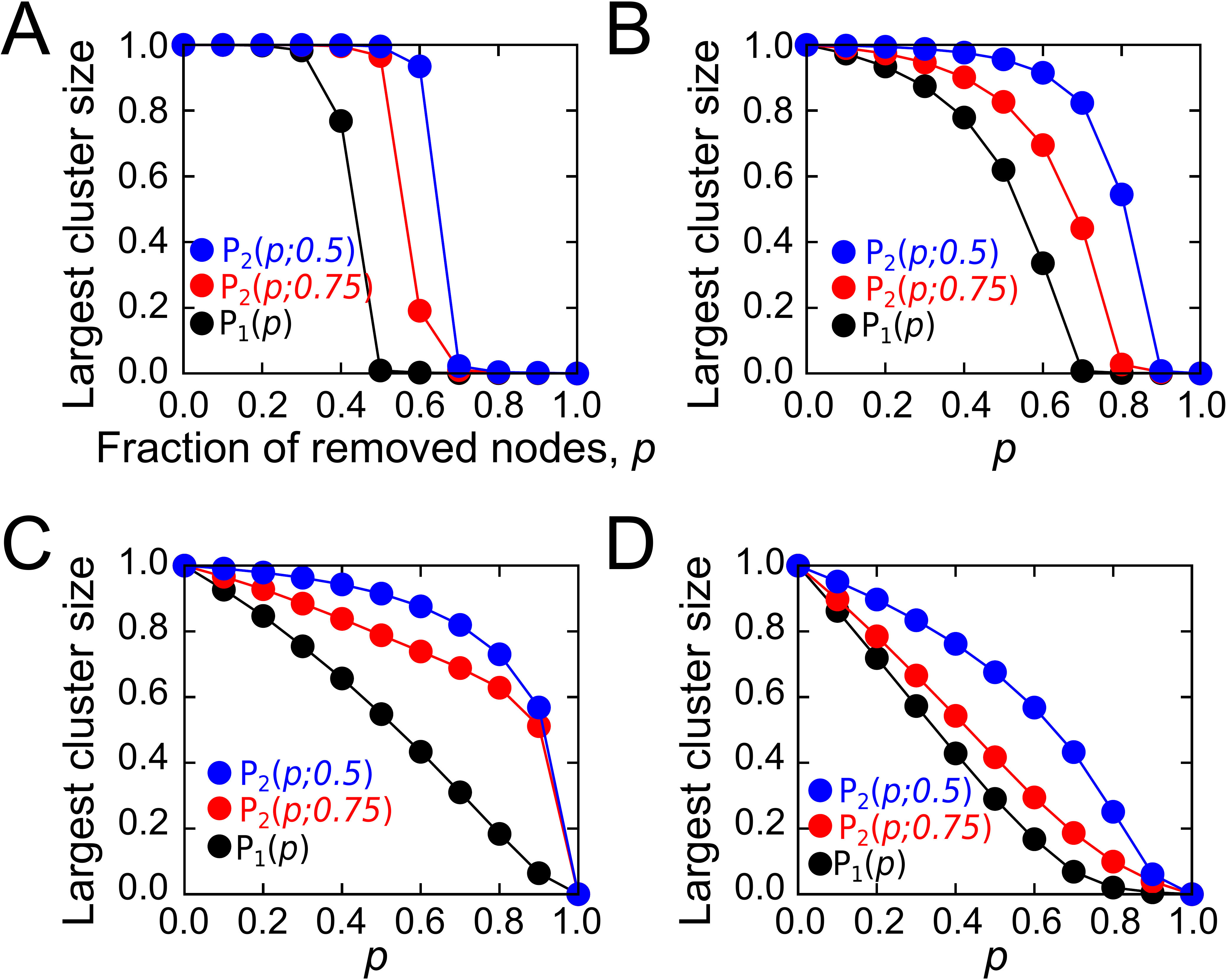}
\caption{\label{fig3} (Color online) Fraction of surviving nodes in the largest cluster as a function
of the removed nodes, $p$. The panels correspond to: (A) Two-dimensional square lattice,
(B) Erdos-Renyi network with average degree $\langle k \rangle =3$, (C) Random
scale-free network with degree exponent $\gamma=2.5$, and (D) Random scale-free network
with degree exponent $\gamma=3$. The network size in all cases was $N=10^5$ nodes.
Black (lower) lines indicate the cluster size $P_1(p)$, after the initial removal of nodes.
The red (middle) lines represent the cluster size after the healing process, $P_2(p;0.75)$,
where a node decides to search for new connections if it has lost at least $q_c=75\%$
of its original neighbors. The blue (upper) line shows the healed cluster size, $P_2(p;0.5)$,
for a corresponding threshold of $q_c=50\%$. The healing distance in all cases is $r_{\rm max}=2$.}
\end{figure}

In Fig.~\ref{fig3} we show the evolution of $P_1(p)$ and $P_2(p)$ as a function of the
removed percentage of nodes, p, in a random removal process. The results for
lattices, Erdos-Renyi, and scale-free networks are well known and we recover
the typical curves for $P_1(p)$ \cite{23}. The self-healing process results in
significantly larger critical points for lattice and ER networks. This delay
indicates that even though a network has been fully disintegrated, our
healing algorithm can reconstitute one large cluster of size $P_2(p)$ among the
surviving nodes. This improvement is obvious even for large threshold
values, such as $q_c$=0.75, when a node delays adding a new link until it has
lost more than 75\% of its neighbors. As we relax this threshold to $q_c$=0.5
the largest cluster reaches higher values over a wider $p$ interval. Lowering
this threshold even more did not produce noticeable improvement. Therefore,
in the following analysis we fix this value to $q_c$=0.5.

Each node seeks new connections within a maximum distance $r_{\rm max}$, and for the
results seen in Fig.~\ref{fig3} we used the minimum possible distance, i.e. $r_{\rm max}$=2.
As we increase the radius of possible new connections the connectivity
becomes easier to achieve for a number of reasons. First, there are more
options because the node can create a link, even if all second-neighbors
have been removed. Second, the effect of long-range shortcuts is known to
favor large-scale connectivity \cite{24}. For example, if the attack results in
small, isolated clusters, then short-range shortcuts will tend to remain
within the same cluster and cannot assist in bridging between different
clusters. On the other hand, long-range shortcuts will have a much higher
probability of bridging otherwise unconnected parts of the network because
they can choose nodes that are far from the immediate neighborhood of their
initiator \cite{25}. The problem with this approach, of course, is that the
associated cost of long links may become significantly higher. It is
important, therefore, to know if the minimum possible cost is enough to
restore the network structure.

\subsection{Efficient Cost-Constrained Repair}

As we will demonstrate below, in the absence of cost restrictions, higher values of $r_{\rm max}$ improve the
repair results, with the optimum case being that $r_{\rm max}$ is unrestricted and
the node can randomly choose any other surviving node in the system. This is a trivial theoretical result,
because long-range links can connect isolated clusters which are far from each other in case of extended damage.
Importantly, however, beyond this trivial result, we show that the minimum
value of $r_{\rm max}$=2 is already sufficient to restore network structure. We
therefore focus our discussion by reporting only results for this worst-case
scenario.

\begin{figure}
\includegraphics[width=.48\textwidth]{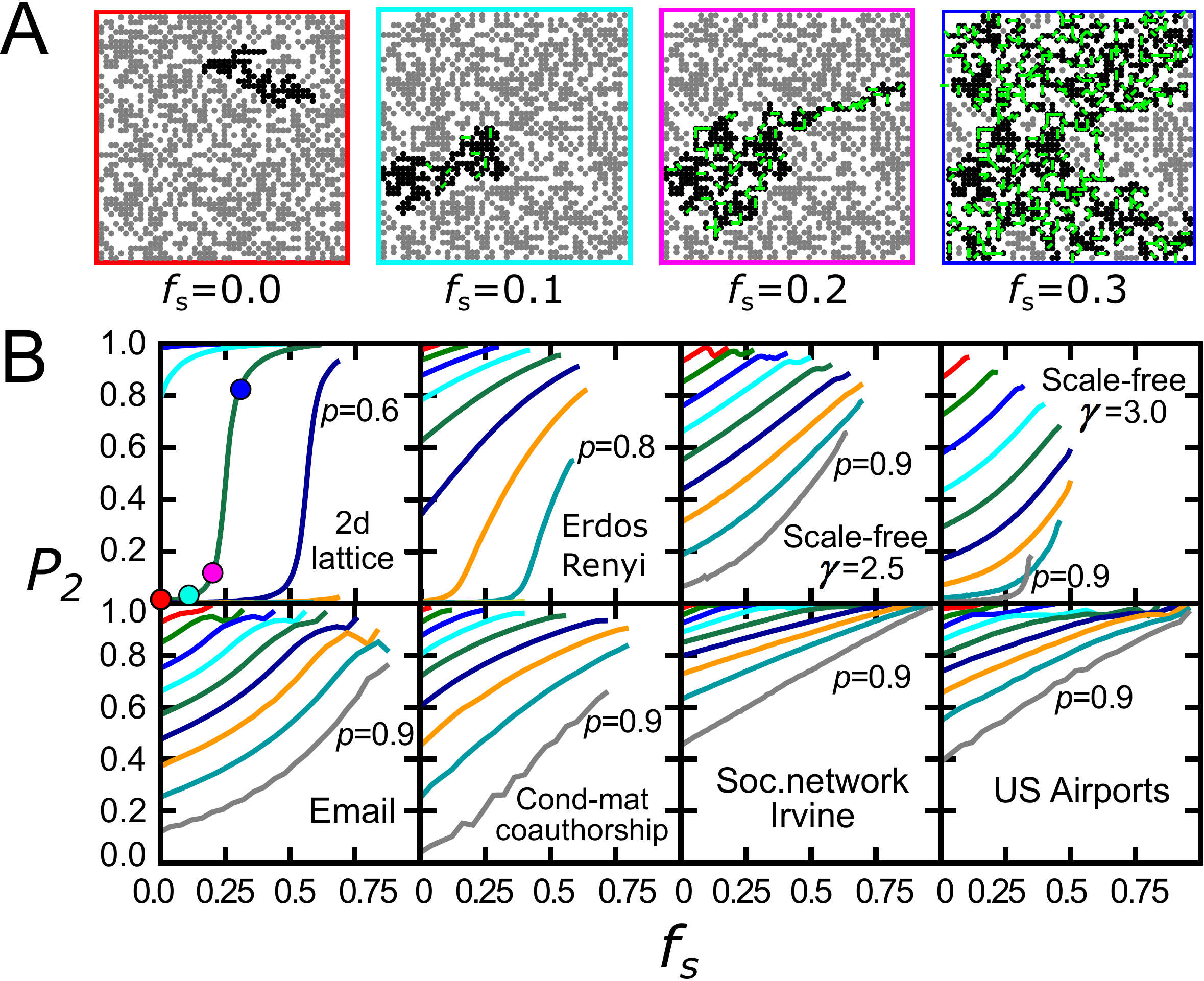}
\caption{\label{fig4} (Color online) (A) Snapshots of the healing process in a square two-dimensional lattice,
after removing $p=50\%$ of the nodes. The nodes that survived the attack are shown in gray color,
with black nodes indicating the largest cluster after a fraction, $f_s$, of the surviving nodes
have generated a new link. The healing links are shown in green (light gray on black nodes). (B) Evolution of the largest cluster
size during healing, as a function of the percentage of surviving nodes, $f_s$, that establish a new link.
From top to bottom, the fraction of removed nodes increases in steps of 0.1, from $p=0.1$ to $p=0.9$ (or to the
value indicated on the plot). The starting point of each curve at $f_s=0$ corresponds to the largest cluster size immediately after the removal of $p$
nodes. The end point of each curve indicates the final cluster size after the end of the healing process.}
\end{figure}

In Fig.~\ref{fig4} we present many examples of the self-healing process, for both
model and real-world networks. A lattice structure offers the simplest
example of self-healing. In Fig.~\ref{fig4}A we present a toy example, in a lattice
where $p=50\%$ of the nodes have been removed. As we add the links to a
fraction fs of the nodes that have lost more than half of their neighbors,
the size of the largest cluster increases and presents essentially a typical
second order percolation transition. This behavior is verified in the first
panel of Fig.~\ref{fig4}B, where the large-scale connectivity is easily restored for
values of $p\leq 0.6$. The second-order transition is observed in all the model
networks, but only when the values of $p$ reach close to their critical values
pc. For $p<p_c$, the initial largest cluster at $f_s=0$ extends over a non-zero
fraction of the remaining nodes and, notably, there is no transition in
these cases. The linear increase of the largest cluster size indicates that
the new links attach small clusters to the spanning cluster, and there is
never any other cluster of significant size. The absence of a transition is
observed in all real networks, even when the initial largest cluster is very
small. For instance, at $p=0.9$, the largest cluster size of the co-authorship
network starts at less than $P_1=5\%$ and with healing it increases almost
linearly up to $P_2~70\%$. Notice that in these plots the process does not
necessarily go up to $f_s=1$, since not all nodes need, or can find, new
connections according to the rules of the algorithm.

\begin{figure}
\includegraphics[width=.48\textwidth]{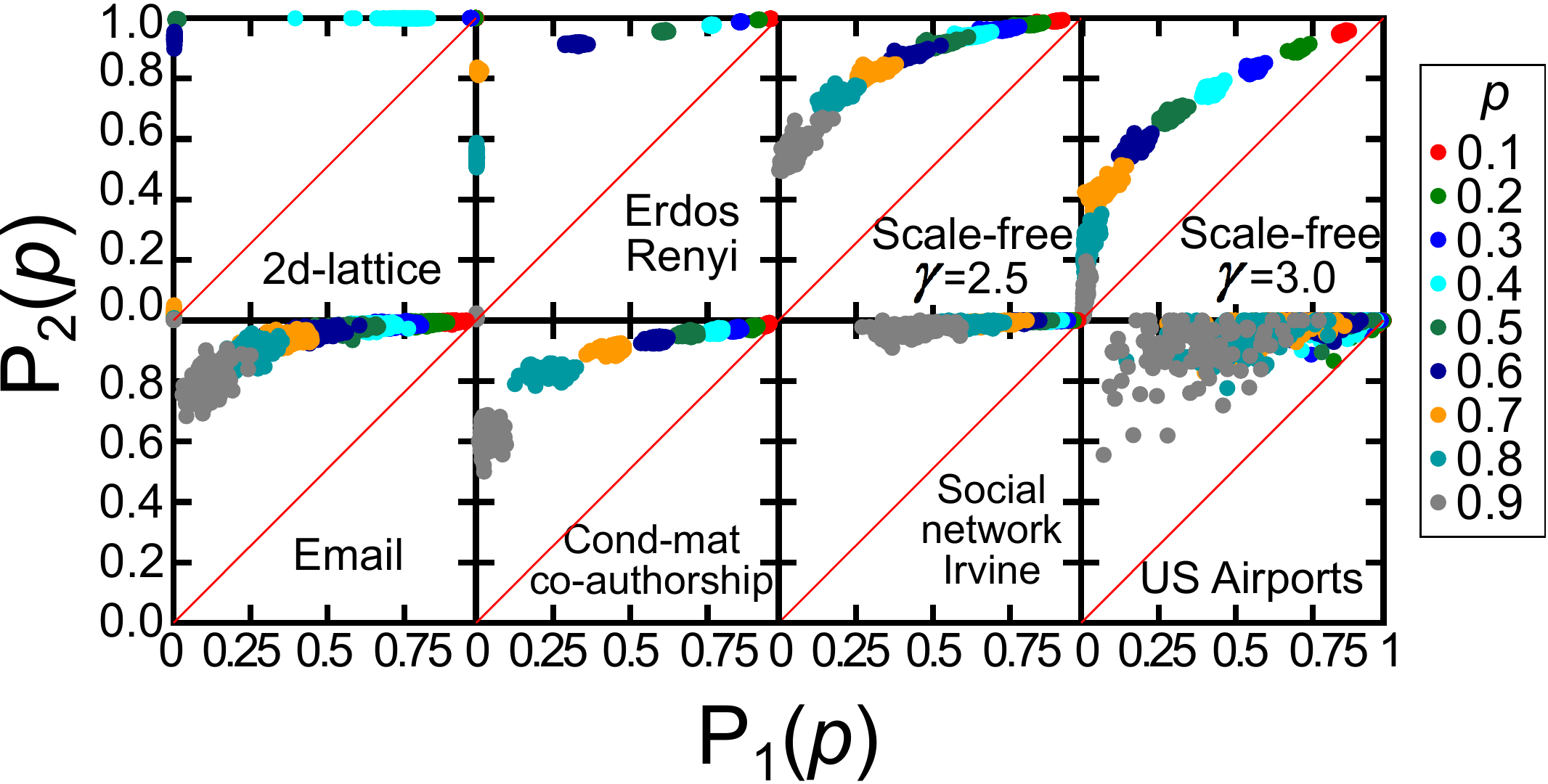}
\caption{\label{fig5} (Color online) How the largest cluster size changes after healing, $P_2(p)$,
as a function of the largest cluster size before healing, $P_1(p)$. Results are
presented for model and real-world networks. The percentage of nodes removed is
represented by the color of a node, increasing from right to left from $p=0.1$ to $p=0.9$. Each scattered point corresponds to a different
realization of the process in each network. The efficiency of the healing process
is demonstrated by the fact that all the points tend to be far from the diagonal,
which would indicate no improvement, and remain close to the maximum efficiency value of $P_2(p)=1$.}
\end{figure}

We then compare the value of the cluster size at the end of healing, $P_2$,
with the cluster size before healing, $P_1$, for different values of $p$ (Fig.~\ref{fig5}).
In lattices, we know that close to the threshold $p\sim 0.5$ the system is in
the critical phase so that a small number of links are enough to restore
connectivity. However, in the plot we can see the large improvement that
healing brings to the tolerance of the system. Even though the network
disintegrates very rapidly as we remove more nodes and increase $p$ to $p\sim 0.6$,
the largest cluster after self-healing practically includes all remaining
nodes, so that $P_2\sim 0.95$. Moving from $p=0.6$ to $p=0.7$ causes a much greater
damage which cannot be restored via our self-healing process and $P_2$ abruptly
drops to $P_2\sim 0$. In practice, the self-healing algorithm has managed to delay
the location of the critical point from $p\sim 0.5$ to $p\sim 0.7$. This simple behavior
of either a fully connected or fully disconnected cluster is rather
idiosyncratic of the highly organized configuration of nodes in the lattice,
which is not found in random models and real-world networks, as we show
below.

The Erdos-Renyi network (with average degree $\langle k \rangle =3$) follows a similar
pattern where removing 70\% of the nodes results to isolated clusters,
$P_1(p)\sim 0$. After healing with our algorithm the largest cluster size
encompasses a large part of the remaining nodes, $P_2(p)\sim 0.8$. This striking
difference indicates that the remaining clusters remain close to criticality
and relatively close to each other, since they can be merged when nodes add
new local links with $r_{\rm max}$=2. Similar improvement is found for the case of
scale-free networks with different degree exponents.

\subsection{Model vs Real-World Networks}

An important difference between random model networks and real-world
networks is usually the level of organization. For instance, typical
empirical networks exhibit higher degrees of clustering and modularity,
features that are missing in the randomly built networks \cite{26}. Nevertheless,
this organization can potentially drastically impact the outcome of the
healing process. In the examples that we studied (Fig.~\ref{fig5}) we see that
self-healing is equally, if not more, efficient in restoring the largest
cluster in empirical networks. In certain cases, e.g. the friendship social
network in Irvine or the email exchange network, practically all the
surviving nodes belong to the same cluster after healing, even though the
starting network state is highly disintegrated, i.e. $P_1(p)<0.2$. Of
particular importance is the result for the network of airport connections,
because it is a spatially embedded structure and we can approximate spatial
distance with network distance, at least to demonstrate the principle that
longer jumps may be very costly to implement. The example highlights the
importance of small $r_{\rm max}$ values. If an airport closes down then the traffic
needs to be redirected to another airport in the general area, but it is
impractical if this distance remains unrestricted. Additionally, in this
example connectivity is crucial for the network to be functional (one needs
to be able to reach any destination). Under the conditions of our
self-healing algorithm, the large-scale connectivity was almost certain in
all our simulations, with more than 80\% of the nodes belonging in the
largest cluster. The large fluctuations, especially in the extreme case
where 90\% of the nodes have been removed, are the result of the small
network size, which includes 332 nodes. This demonstrates that healing may
become unpredictable for very small networks, where one or two links may be
enough to connect the small number of clusters.

\begin{figure}
\includegraphics[width=.48\textwidth]{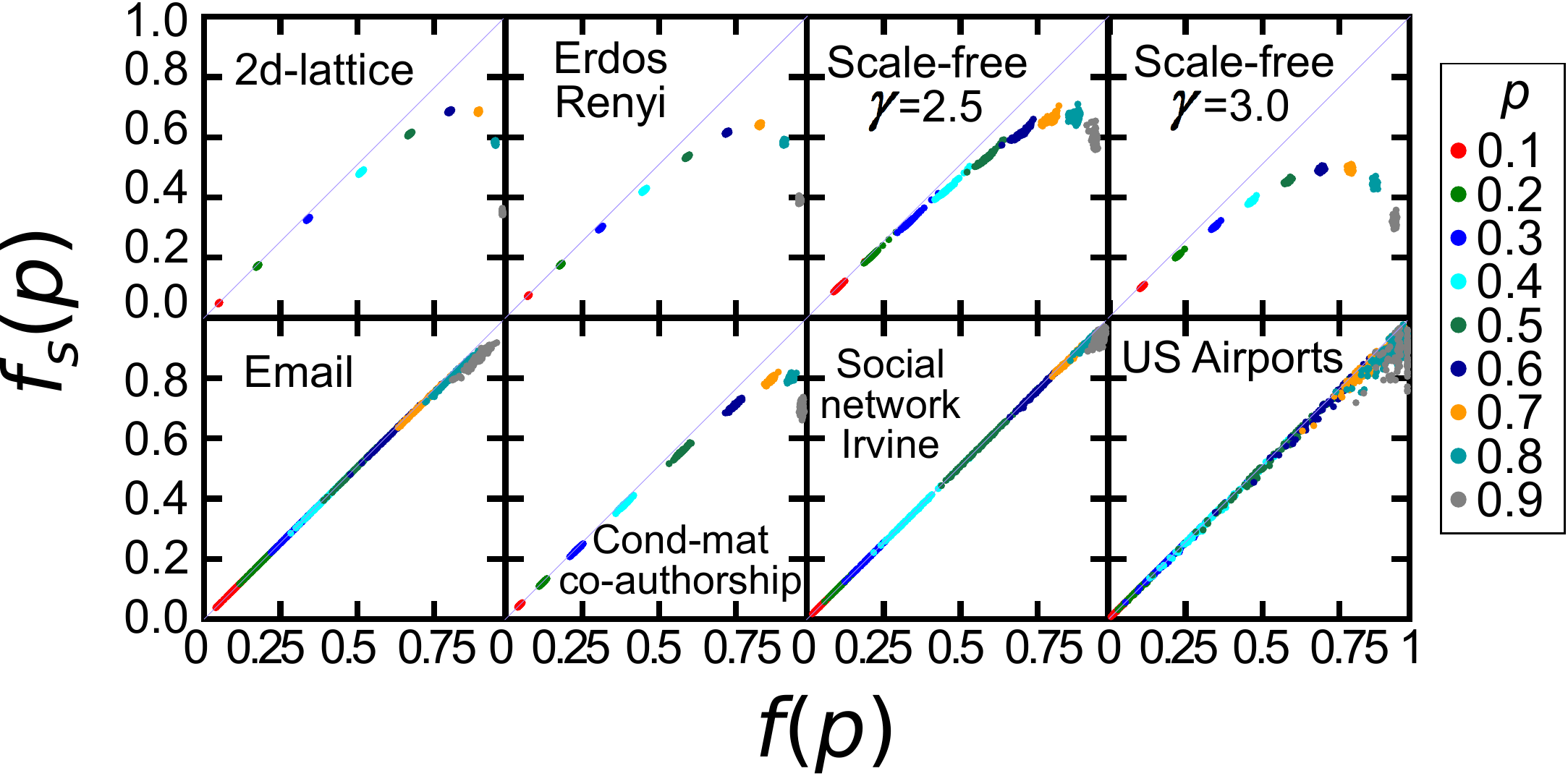}
\caption{\label{fig6} (Color online) Fraction of the surviving nodes that successfully established new connections,
$f_s$, as a function of the fraction of nodes, $f$, that attempted to establish new connections
following the removal of a percentage $p$ of nodes. The value of $p$ increases from left to right.
The points that are closer to the diagonal indicate a higher percentage of success.}
\end{figure}

The percentage of nodes that find themselves with fewer neighbors than the
threshold increases with the number of nodes that are removed. These nodes
do not necessarily manage to find a new neighbor within the distance $r_{\rm max}$=2,
either because all these nodes have been removed or because there is already
a connection. As we show in Fig.~\ref{fig6}, all nodes in lattice and in model random
networks manage to find a new neighbor as long as less than 50\% of the
initial nodes have been removed. As the extent of damage increases, a larger
percentage of nodes need new connections, but less than half of those manage
to do that. On the contrary, in real networks all nodes manage to find new
connections almost independently of the extent of damage in the network.

\begin{figure}
\includegraphics[width=.48\textwidth]{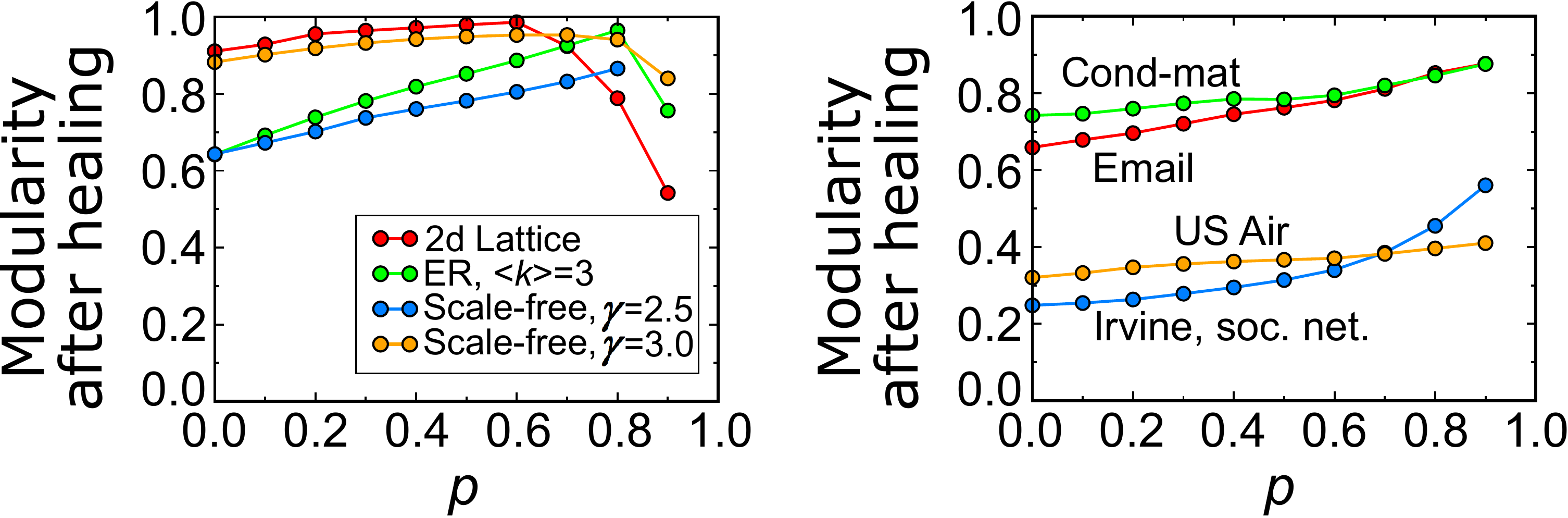}
\caption{\label{fig7} (Color online) Effect of the self-healing process on the network
modularity, as a function of the removed nodes. The modularity at $p=0$ corresponds
to the original network modularity. For $p>0$, we consider the modularity of the
network after healing with $r_{\rm max}=2$.}
\end{figure}

Even though functionality is restored, since the majority of nodes can now
communicate with each other through the structural repair that reconnected a
large cluster, it is not clear what impact restoration of this particular
structural feature may have on other topological/structural features. We
therefore study the effect on the modularity of the healed network compared
to the modularity of the undamaged network as an example of a potentially
important structural feature which is not necessarily correlated to the size
of the largest cluster. We use the standard definition for modularity $Q$, as
the fraction of links between modules, minus the expected number of links
within these modules for a random graph with the same node degree
distribution \cite{26}. For Fig.~\ref{fig6} we report the maximum possible value of $Q$,
using the algorithm from Ref. \cite{27}. In Fig.~\ref{fig7} we show changes in modularity
after a random removal of $pN$ nodes and the application of the self-healing
algorithm. A common trend in both model and real networks is that modularity
increases considerably with increasing $p$. This shows that the healing effect
tends to create significantly stronger modules, and nodes tend to become
more connected within the same network area. Of course, this is the result
of the short-range links in the presented simulations, where $r_{\rm max}$=2. In
practice, the healing algorithm replaces the links that are removed with
local links, enhancing thus the modular character of the network. The only
instances where modularity decreases are at large values of $p$ in the model
networks. This result can be explained by Fig.~\ref{fig6} where only a small
percentage of nodes manage to find new connections and therefore the form of
the resulting structure is not very different than the damaged structure.

\subsection{Self-Healing vs Random Repair}

The efficiency of our method can be tested against a random insertion of
links under a similar set of constraints, as for the self-healing algorithm.
For example, we can add the same
number of links at the same distance, but random nodes are chosen to create
new links, rather than those who have lost most of their neighbors. In this
way, we can detect if the crucial factor is the number of edges or the
particular features of the nodes that select to establish new connections.
Even if our model results did not exceed those of random insertion, our
method is based on a self-organized algorithm that provides a simple local
decision scheme, where there is no need for a central authority to
coordinate actions around the network. Still, we demonstrate below that our
method, which we tested in a number of model and real-world networks,
significantly exceeds the results of random decisions.

\begin{figure}
\includegraphics[width=.48\textwidth]{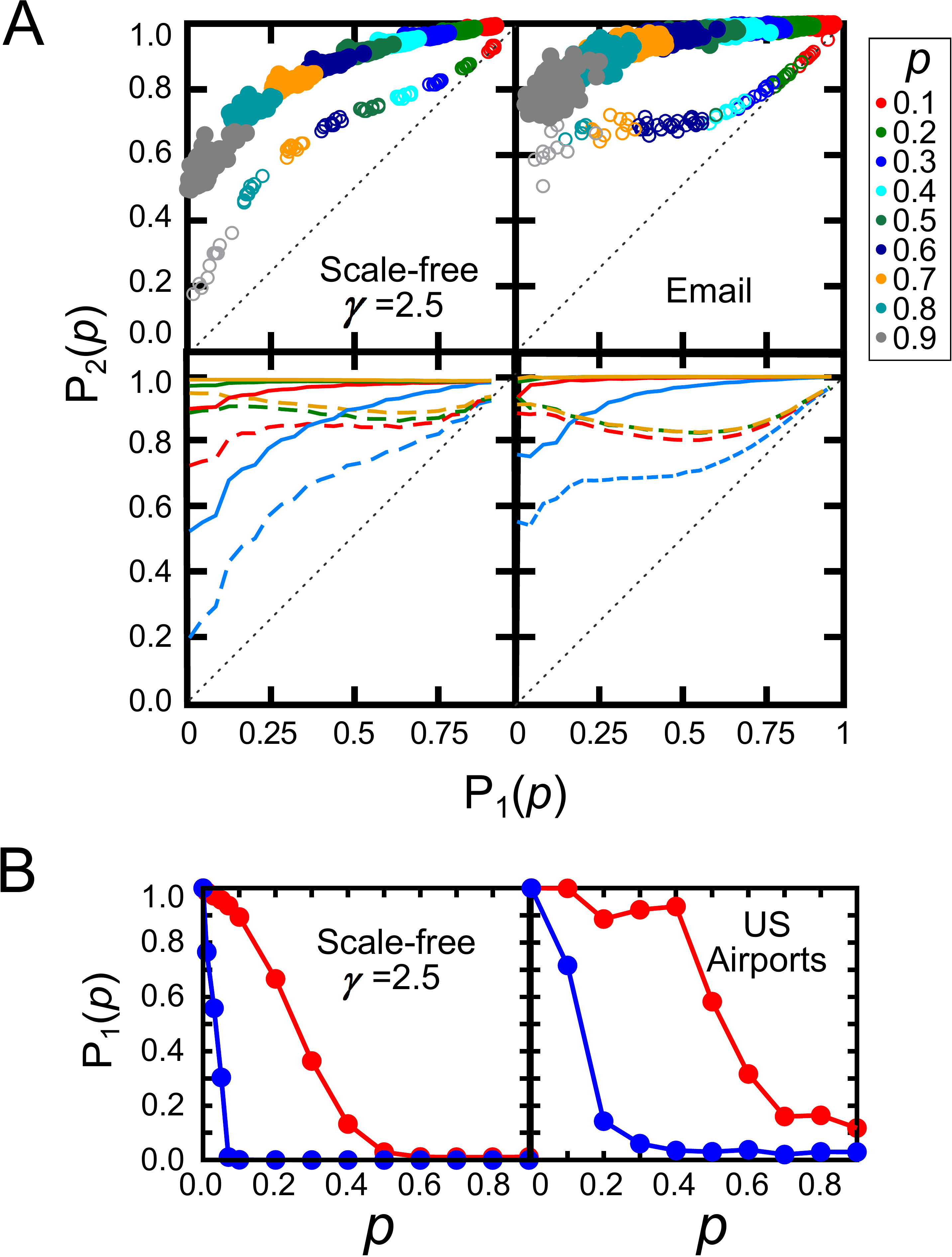}
\caption{\label{fig8} 
(A) Top row: Comparison of the result for the
self-healing algorithm (filled symbols) with adding the same number of links starting
from random nodes (empty symbols). In both cases, $r_{\rm max}=2$.
Bottom row: Comparison of the self-healing algorithm (continuous lines) with the random
case (dashed lines), for different $r_{\rm max}$ values. From bottom to top: $r_{\rm max}$=2,
3, 4, and $\infty$.
(B) Comparison of the largest cluster size after an intentional attack.}
\end{figure}

We assess the effectiveness of the algorithm by comparing our results to such
a null model. In this way, we can
determine if there is a benefit to choosing which nodes create new links
based on $q_c$. For each realization of nodes removal with the self-healing
algorithm we calculated how many new links were introduced in the system,
resulting to $P_2(p)$. We then added the same number of links to the damaged
network, but this time the link originated from a random node instead of the
node that had lost more than a fraction $q_c=50\%$ of its neighbors. These new links
were established within a distance $r_{\rm max}$=2, so that the total cost remained
the same in both cases. The result in the top row of Fig.~\ref{fig8}A shows that the self-healing
algorithm performed better in all cases compared to this null model. The
reason is that when a random node decides to establish a new link, it is
possible that many of its neighbors still survive and the new link does not
bring together isolated parts of the network. This is also why when the
removed percentage of nodes is small, e.g. 10-20\%, there is no increase in
the largest cluster size, unless we use the nodes that have lost most of
their neighbors. 

We also studied the effect of $r_{\rm max}$ on the results (Fig.~\ref{fig8}).
As expected, we find that as we increase the maximum distance for new connections,
the repair process becomes more efficient. As shown in the plot for the self-healing algorithm, $r_{\rm max}=3$
is already enough to completely restore connectivity among all the remaining nodes,
practically under any conditions. The optimal case is found for $r_{\rm max}\geq 4$. The results for the 
random model are consistently inferior to those of our algorithm. The trend of better healing with
increasing $r_{\rm max}$ remains, but now even for unrestricted $r_{\rm max}=\infty$, the value
of the repaired largest cluster cannot reach the optimal value, $P_2(p)=1$.

\subsection{Random vs Targeted Attack}

Until now, we studied the healing process on networks where the nodes were
randomly selected for removal. Another important removal technique simulates
intentional attack on the network, where nodes are removed in descending
order of their degree, i.e. the highest connected nodes are removed first.
This attack requires a much smaller percentage of removed nodes to destroy
the network, since all the hubs that glue the network together are removed
and the network disintegrates into small pieces. The self-healing algorithm
is efficient in this case, too (Fig.~\ref{fig8}B). The critical value of $p$ for a
scale-free network with degree exponent $\gamma$=2.5 moves from $p\sim0.05$ to $p\sim 0.5$.
This is a huge difference that allows the network to operate even after
losing the most connected nodes in its backbone. Similarly, in the case of
the airport connection network, the removal of the hubs leads to a
disconnected network when 20\% of the nodes are removed, but the healed
network manages to preserve the largest cluster even if more than 60\% of the
nodes are removed.

\section{Discussion}

In this paper we make explicit a fundamental feature of damage in complex
networks and use it to design a very simple self-healing algorithm to
restore connectivity-based function. We show that even just this simple
example is enough to restore large-scale connectivity in a severely damaged
network, while keeping the reconnection cost low. In this particular
scenario of structure enabling function, a node decides based on its own
link losses to add a new link to one of its second neighbors (provided at
least one survives) without considering any other parameter. This approach
has the added benefit that it is strictly local, and every node can make the
decision autonomously. The cost is also lower with this strategy compared to
selecting random nodes for the same number of links.

The critical insight that working to restore function may not rely on
restoring the initial structure allowed us to design a threshold, $q_c$. This
threshold acts as a natural filter to direct the addition of the links
towards the areas where function has been most affected by the damage in
this connectivity-based scenario. 

In fact, this strategy represents a worse-case scenario, because the nodes
make new connections only after the removal of all nodes, or equivalently if
the attack takes place faster than the nodes can react. In an alternate
version, the algorithm could be applied in a dynamic fashion so that a node
continuously monitors its neighbors and decides to add new links based on
current information, i.e. as soon as it notices that the number of its
neighbors falls below the threshold. Under this scenario, a node could
establish more than one new links during the process and it would be easier
to preserve long-range connectivity, simply because of the addition of a
larger number of links. What we showed above is that this extension is not
necessary since a similar result can be achieved with much fewer new
connections.

A few interesting parallels can be drawn between the self-healing algorithm
and Achlioptas processes \cite{28}. In the latter, we start from an empty network
of $N$ nodes which grows by adding new links according to the process rules.
These rules can either favor or discourage the emergence of a largest
cluster. In our case, the starting point is an already fractioned network,
and the goal is to merge all clusters into one giant network with as few
links as possible. A key difference is that we only use local decisions,
while in typical Achlioptas processes the information of the involved
cluster sizes is required. 

\begin{acknowledgments}
We thank the Dept.~of Homeland Security for funds in support of this
research through the CCICADA Center at Rutgers, and NSF EaSM grant No.~1049088.
\end{acknowledgments}

\section{APPENDIX: DATASETS}

We analyzed four different networks, based on the following datasets:

\begin{enumerate}

\item Email network: Email messages sent at the Computer Sciences
Department of London's Global University
\url{http://lisgi1.engr.ccny.cuny.edu/~makse/SOCIAL/Emailcontacts.dat.gz}.

\item Cond-mat co-authorship: The network of co-authorship in preprints
submitted to the cond-mat section of arxiv.org \cite{29}.

\item Irvine social network: The dataset was downloaded from
\url{http://toreopsahl.com/datasets/#online_social_network} and has been analyzed
in \cite{30}. It includes online messages sent among students at the University
of California, Irvine, through a Facebook-like Social Network.

\item USA airport network: A link indicates that two networks are
connected by a direct flight. This dataset refers to flights in 1997 and can
be downloaded at
\url{http://vlado.fmf.uni-lj.si/pub/networks/data/mix/USAir97.net}.
\end{enumerate}

\end{document}